\documentclass[12pt]{article}
\usepackage{amsmath}
\usepackage{amssymb}
\usepackage{graphicx}
\input epsf
\usepackage{dcolumn}
\usepackage[pdftex,
            pdfauthor={A.V. Yurov, V.A. Yurov},
            pdftitle={The Landau-Lifshitz equation, the NLS, and the magnetic rogue wave as a by-product of two colliding regular positons},
            pdfsubject={},
            pdfkeywords={Magnetic Rogue Wave; Spine waves; Landau-Lifshitz-Glbert equation; Nonlinear Schrödinger equation; Darboux transformation; Positons},
            pdfproducer={Latex with hyperref},
            pdfcreator={pdflatex}]{hyperref}
\hypersetup{colorlinks=true, citecolor=blue, linkcolor=red}

\newcommand{\beq}{\begin{equation}}
\newcommand{\beqn}{\begin{equation*}}
\newcommand{\enq}{\end{equation}}
\newcommand{\enqn}{\end{equation*}}
\newcommand{\eb}{{\rm e}}

\newcommand{\R}{{\mathbb R}}
\newcommand{\N}{{\mathbb N}}

\renewcommand{\Re}{\text{\rm Re}}
\renewcommand{\Im}{\text{\rm Im}}
\renewcommand{\l}{\lambda}

\newcommand{\footremember}[2]{%
    \footnote{#2}
    \newcounter{#1}
    \setcounter{#1}{\value{footnote}}%
}
\newcommand{\footrecall}[1]{%
    \footnotemark[\value{#1}]%
}
\begin{document}
\font\bss=cmr12 scaled\magstep 0
\title{The Landau-Lifshitz equation, the NLS, and the magnetic rogue wave as a by-product of two colliding regular ``positons''}
\author{A. V. Yurov \footremember{Kant}{Immanuel Kant Baltic Federal University, Institute of Physics, Mathematics and Informational Technology,
 Al.Nevsky St. 14, Kaliningrad, 236041, Russia} \footnote{AIUrov@kantiana.ru}
 \and V. A. Yurov\footrecall{Kant} \footnote{vayt37@gmail.com}}%
\date {}
\renewcommand{\abstractname}{\small Abstract}
\maketitle
\begin{abstract}
In this article we present a new method for construction of exact solutions of the Landau-Lifshitz-Gilbert equation (LLG) for ferromagnetic nanowires. The method is based on the established relationship between the LLG and the nonlinear Schr\"odinger equation (NLS), and is aimed at resolving an old problem: how to produce multiple-rogue wave solutions of NLS using just the Darboux-type transformations. The solutions of this type –- known as P-breathers -- have been proven to exist by Dubard and Matveev, but their technique heavily relied on using the solutions of yet another nonlinear equation, Kadomtsev-Petviashvili I equation (KP-I), and its relationship with NLS. We have shown that in fact one doesn't have to use KP-I but can instead reach the same results just with NLS solutions, but only if they are dressed via the {\em binary} Darboux transformation. In particular, our approach allows to construct all the Dubard-Matveev P-breathers. Furthermore, the new method can lead to some completely new, previously unknown solutions. One particular solution that we have constructed describes two ``positon''-like waves, colliding with each other and in the process producing a new, short-lived rogue wave. We called this unusual solution (rogue wave begotten after the impact of two solitons) the ``impacton''.
\end{abstract}
\thispagestyle{empty}
\medskip
\section{Introduction} \label{sec:Intro}
The Darboux transformation (DT) \cite{D1882}, \cite{D1972} for one-dimensional Schr\"odinger equation serves as an important case of so-called isospectral symmetries, which are directly related to the factorisation method \cite{Infeld} and are used to construct the ``integrable'' potentials from pre-existing ones. However, the DT also possesses one very particular feature that distinguishes it from most other isospectral symmetries-related transformations, namely: just as it produces the new potentials it also allows to find and control the eigenvalues of the corresponding Hamiltonians. For example, after a single DT has produced a new non-singular potential from an old (also non-singular) one, we invariably end up with a new Hamiltonian, whose spectrum will either be completely identical to the original or differ from it by just one (lowest) eigenvalue. It is this property that makes it possible to realize the supersymmetric (double degeneracy) \cite{Schulze} and parasupersymmetric (triple degeneracy, see \cite{paraS1990-1}, \cite{paraS1990-2} and \cite{YYY}) models. Furthermore, the DT paves a unified pathway to many famous algebraic methods of integration, including (but not limited to) the formalism of second quantization and the ladder operators, that arise during the construction of irreducible representations of symmetry group $R_3$, etc. Consequently, we end up wielding a simple yet robust method uniquely designed to generate important applied relationships between various special functions (the Hermitian and Legendre polynomials, the hypergeometric function, etc).

It would not be much of an overstretch to claim that for all reputed examples of the potentials for whom the Schr\"odinger equation can actually be analytically solved, all of them can be generated by DT in one of two possible ways. First, one can simply apply DT to the already known integrable potential (for example, to zero-valued potential, as in the case of the reflectionless soliton potentials). Alternatively, one can combine a chain of DT's with an additional invariance condition which shall guarantee the preservation of the potential's form. Interestingly, the derivations for the latter method can be significantly simplified if instead of a standard DT one would make use of dressing chains of discrete symmetries  \cite{VS1993}. In this case adding the closure condition on the dressing chains with $N$ ``links'' results in a unified technique for generation of the integrable potentials, ranging from the harmonic oscillator and up to the potentials defined via the Painlev\'e transcendents (and their generalizations).

Another general reason for a wide applicability of DT stems from the fact that {\em any} linear PDE of a form
\begin{equation}
\frac{\partial \psi}{\partial t}=\sum_{k=1}^N
A_k(x,t)\frac{\partial^k\psi}{\partial x^k}, \label{ffirstt}
\end{equation}
where $\psi=\psi(x,t)$ and $A_k(x,t)$ are matrix-valued functions allows for the DT-type discrete symmetries. In turn, a series of physically important nonlinear differential equations can be reduced to a system of equations, at least one of whom will be of a form (\ref{ffirstt}). In particular, such famous nonlinear integrable equations as Korteweg-de Vries (KdV) equation, the Davey-Stewartson equation (DSE) and the nonlinear Schr\"odinger equation (NLS -- we'll say lots more on it a bit later) are reducible to completely linear systems of equations called the Lax pairs.

Among the possible applications of DT one of the most important is the construction of exact solutions for the soliton equations that accounts for (among other things) the interaction with an arbitrary background solution (noise) of a nonlinear system. The versatility of DT for this particular purpose is due to the fact that to find new solutions for a nonlinear problem one only has to know a few partial solutions of a system of linear differential equations, with the aforementioned nonlinear problem serving as a compatibility condition for this system. This feature of DT allows to both account for all possible free parameters, and construct such new solutions that are almost unobtainable via other methods (such as inverse scattering method etc). For example, a simple addition of a differentiation w.r.t. the spectral parameter to the standard DT algorithm has allowed V. B. Matveev to discover a completely new type of solutions called {\em positons} for such well-known and well-studied model as KdV equation \cite{Matveev1}.

Furthermore, an apt application of the discrete symmetries chains formalism to the DT has helped to uncover a multiplicity of unexpected symmetries between the familiar integrable hierarchies. For instance, the hierarchies of KdV, mKdV, the exponential Calogero–Degasperis equation (CDE) and of the elliptic CDE ends up connected with each other via the dressing chains of discrete symmetries  \cite{Bor-Zyk1998}. A similar relationship has been shown to exist between the NLS and the Toda chains (only this time with the aid of another isospectral symmetry -- the Schlesinger transformation) \cite{Shabat1999}, \cite{Yurov2003}.

In this article we will discuss and make use of a certain generalization of the DT -- the {\em binary Darboux transformation}; its purpose would lie in the construction of a special class of exact solutions of nonlinear Schr\"odinger equation -- the so-called ``{\em rogue waves}''. We will also show that utilization of these discrete symmetries is sufficient for a production of all known P-breather solutions and will even provide a completely new solution -- the ``{\em impacton}'': a short-lived rogue wave generated by the collision of two nonsingular positons.

But first thing first. Let us begin by discussing a particular model we will consider: a magnetization of ferromagnetic nanowire.

One of the primary tools essential for studying the magnetic properties of ferromagnetic solids is the famous Landau-Lifshitz-Gilbert (LLG) equation \cite{LL}, \cite{Gilbert}. Its purpose lies in describing the dynamics of a magnetization vector field $\vec M$ of the ferromagnetic in response to an effective magnetic field $\vec H_\text{eff}$ (which includes an external magnetic field, and the effects of anisotropy and of spin exchange interaction) and an electric current $j_e$\footnote{It is also possible to account for some additional effects such as a presence of magnetic defects right in \eqref{LLG_equation} by simply adding the required terms to the effective field $\vec H_\text{eff}$ (see, for example, \cite{Visintin}).} and have a form ($t$ being the time variable):
\beq \label{LLG_equation}
\frac{\partial \vec M}{\partial t} = - \gamma \vec M \times \vec H_\text{eff} + \frac{\alpha}{M_s} \vec M \times \frac{\partial \vec M}{\partial t} + \vec F_j,
\enq
where $\gamma$ is the gyromagnetic ratio, $\alpha$ is called the Gilbert dampening parameter, $M_s$ is the saturation magnetization and $\vec F_{j}$ is called the spin-transfer torque. The equation describes the dynamics of the field $\vec M$ as a precession around the field $\vec H_\text{eff}$ (the first term), combined with gradual dissipation of all components of $\vec M$ orthogonal to $\vec H_\text{eff}$ (the second term, which acts on $\vec M$ as a sort of a ``viscous force''). The last term accounts for the effect of polarization $P$ of an electric current $j$ that goes through the ferromagnetic, and for the torque this polarization inflicts on the overall magnetization.

In order to solve \eqref{LLG_equation} one has to first define the geometry of the problem as well as the relative strength of parameters $\gamma$ and $\alpha$, followed by subsequent rewriting of the vector equation as a system of scalar differential equations. Most astonishingly, many of those geometries lead to one very specific and famous nonlinear equation: the {\em nonlinear Schr\"odinger equation} (NLS). This is not a coincidence, as the equivalence between the LLG and NLS has already been demonstrated as early as in 1970s \cite{Lakshmanan}, \cite{ZT}, and this connection has been extended for some generalized NLS models as well \cite{Kundu}. Just as a few particular examples, the LLG written for ferromagnetic nanowires (with no damping) reduces to straightforward NLS \cite{ZLLWFL}; magnetization of a one-dimensional multicomponent magnonic crystal is shown to be governed by NLS with spatially-dependent coefficients \cite{GSD}; finally, the flat multilayered ferromagnetic crystal whose easy axis is normal to the layers can also be reduced to a standard NLS with a potential containing a sequence of $\delta$-functions \cite{GGG}.

In this article we will study a simpler of these cases: the ferromagnetic nanowire. For simplicity, we will assume that the nanowire is parallel to $x$-axis, and that its anisotropic term and external field $H_\text{ext}$ has only $z$ components (i.e. that we are working with a perpendicular anisotropic ferromagnetic nanowire), just like the one described in \cite{ZLLWFL}. In this case the spin exchange interaction will produce the term proportional to $\vec M_{xx}$, the anisotropy (with good accuracy) will be proportional to $M_z$ ($z$-component of $\vec M$), and the current along the nanowire will produce the spin-transfer torque proportional to $\vec M_x$:
\beq \label{LLG-nanowire1}
\frac{\partial \vec M}{\partial t} = - \gamma \vec M \times \left(\frac{2K}{M_s^2} \frac{\partial^2 \vec M}{dx^2} + \left(\frac{K}{M_s}-4\pi\right) M_z \vec k + H_\text{ext} \vec k\right) + \frac{\alpha}{M_s} \vec M \times \frac{\partial \vec M}{\partial t} + \frac{P j \mu_B}{e M_s} \frac{\partial \vec M}{\partial x},
\enq
where $\vec k$ is a $z$-axis unit vector, $B$ is the exchange constant, $K$ is the anisotropy coefficient, $\mu_B$ is the Bohr magneton and $e$ is the electron charge. The first thing we notice here is that the equation \eqref{LLG-nanowire1} can be simplified by a simple rescaling $\vec M \to \vec m$, $t \to \tau$ and $x \to \xi$:
\beq \label{m_tau_xi}
\vec m = \frac{\vec M}{M_s} , \qquad \tau = \frac{t}{t_0} = t \gamma (K - 4\pi M_s), \qquad \xi = \frac{x}{x_0} = x \sqrt{\frac{K-4\pi M_s}{2B}},
\enq
which produces the following dimensionless equation
\beq \label{LLG-nanowire2}
\frac{\partial \vec m}{\partial \tau} = - \vec m \times \left(\frac{\partial^2 \vec m}{d \xi^2} + (m_z + h_\text{ext})\vec k\right) + \alpha \vec m \times \frac{\partial \vec m}{\partial \tau} + B_j \frac{\partial \vec m}{\partial \xi},
\enq
where we have introduced the following new notation:
\beqn
h_\text{ext} = \frac{H_\text{ext}}{K-4\pi M_s}, \qquad B_j = \frac{P j \mu_B}{e M_s \gamma \sqrt{2B(K-4\pi M_s)}} = \frac{P j \mu_B}{e M_s} \frac{t_0}{x_0}.
\enqn
Next, recalling that $|\vec m| \le 1$, let us introduce a new complex-value function $q(\xi, \tau)$ that is related to the normalized magnetization field as follows:
\beq \label{q}
m_x + i m_y = q, \qquad m_z^2 = 1- |q|^2.
\enq
Finally, let us make a reasonable assumption regarding the magnitude of $|q|^2$: let's assume that $z$ component of magnetization is much stronger than the other two, so $|q|^2 \ll 1$ (the so-called long-wave approximation, \cite{KIK}). This condition implies that we can keep only those nonlinear terms in \eqref{LLG-nanowire2} that are of the magnitude $|q|^2 q$. Furthermore, if the dampening factor is also very small ($\alpha \ll 1$ and does not exceed the maximal value of $q$), we would also be able to omit the dampening term altogether, thus ending up with the following equation:
\beq \label{NLS-1}
i q_{\tau}=q_{\xi\xi} + \frac{1}{2}|q|^2 q+i B_{_J}q_{\xi}-\omega_0 q,
\enq
where $\omega_0 = 1 + h_\text{ext}$. The equation \eqref{NLS-1} is nothing else but a nonlinear Schr\"odinger equation. The last purely cosmetic changes that we can do here is reduce \eqref{NLS-1} to a canonical form by introducing the new variables $\tau \to t'$, $\xi \to x'$:
\beq \label{New-var}
\displaystyle{
\tau=4t', \qquad \xi=2x' - 4 B_{_J} t', \qquad q={\rm e}^{i\omega_0 \tau} {\bar u}(t',x')},
\enq
which allows one to obtain for $u(t,x)$ the standard NLS equation \eqref{NLS}:
\beq \label{NLS-right}
i u_{t'} + u_{x'x'} + 2|u|^2 u = 0,
\enq
where for all further calculations we will omit the accents $'$ and will write just $x$ and $t$.

The equation \eqref{NLS-right} has been under a careful scrutiny ever since its inception, so as a result we now know quite a lot about its possible solutions (``bright'' and ``dark'' solitons, breathers, N-soliton solutions etc.). However, perhaps the most interesting solutions are the so-called rogue wave solutions. First discovered in 1983 by Peregrine \cite{P} and being the first known completely localized regular solution of a form:
\beq \label{PerS}
u(x,t) = \left(1-\frac{4(1 + 4 i t)}{1+4x^2+(4t)^2}\right) \eb^{i t}.
\enq
This solution, called the Peregrine soliton, asymptotically behaves as a simple plane wave, but it also has the exact behaviour of the dreaded oceanic ``rogue waves'', example of which has been detected in the North Sea on the very first day of 1995, inflicting some (thankfully) minor damage to the Draupner platform (the offshore hub for Norway's gas pipelines) \cite{HJT}. In later years, the ``rogue waves'' have become a subject of active explorations and experiments; they have been observed in the optical fibers \cite{Po10}, in the waves generated in the multicomponent plasma \cite{Pp11}, and even in the experimental water tank \cite{Pw11}. So, it was only a matter of time until the attention would turn to the possibility of rogue waves forming more complex structures, possibly consisting of multiple rogue waves. However, until recently all the attempts to construct such solutions have been nothing but exercises in futility, as it soon became clear that the standard techniques -- such as the Darboux transformation (see Sections \ref{sec:Lax} and \ref{sec:Binary_DT} for more information) -- fail to produce any Peregrine soliton-like solutions past the already known ones (the exact reasons for this will be discussed in Sections \ref{sec:Reduction_Restriction} and \ref{sec:Binary_DT}). This vicious cul-de-sac has only been broken in \cite{DGKM}, published in 2010, which proposed a way to actually construct the multi-rogue waves solution by essentially going around the obstacles and working not with NLS, but with another equation, Kadomtsev-Petviashvili I (KP-I), and afterwards using the relationship between NLS and KP-I to get the required NLS solutions. This technique has allowed the authors to construct a set of new solutions, including so-called ``P''-breathers.

However, in this article we will adopt a different approach by sticking to NLS and demonstrating that it is still possible to achieve the same goal using the techniques of Darboux transformation -- provided one uses not a standard, but a {\em binary} Darboux transformation. Furthermore, we will show that our approach not only allows to get the same P-breathers as in \cite{DGKM}, \cite{Dubard_Matveev} but also a previously unknown solution with a very unique properties: it describes a collision of two slowly-moving regular positons (or negatons), which produces a short-lived rogue wave, which we have tentatively called the ``{\em impacton}''. One interesting aspect of the impacton model lies in slowness of movement of its parental solitons which might make this solution a very good candidate for observation on particular ferromagnetic nanowires.


\section{NLS and its zero-curvature condition} \label{sec:Lax}
In this section we begin the exploration of the NLS equation
\beq \label{NLS}
i u_t + u_{xx} + 2|u|^2 u = 0,
\enq
and will attempt to answer the question posed in the introduction: does there exist a simple way to produce a new non-trivial Peregrine-style solution from the already known one?

In order to succeed in our endeavour we will first have to reduce our nonlinear problem to a slightly more manageable linear system. The system in question is the zero curvature condition, also known as the Lax pair for the NLS.
\beq \label{Lax_a}
\begin{split}
\Psi_x & = i \sigma_3 \Psi \Lambda + i U \Psi, \\
\Psi_t & = 2 i \sigma_3 \Psi \Lambda^2 + 2 i U \Psi \Lambda + W \Psi,
\end{split}
\enq
where $\Lambda$ is a $2\times 2$ diagonal complex-valued matrix,  $\sigma_3$ is Hermitian, and is called the third Pauli matrix
\beq \label{sigma_3}
\sigma_3 = \left(\begin{array}{ccc}
1 & 0\\
0 & -1
\end{array}\right),
\enq
$\Psi=\Psi(x,t)$ is a $2 \times 2$ matrix-valued function, and the matrices $U$, $V$ are defined as follows:
\beq
U = \left(\begin{array}{ccc}
0 & \bar u\\
u & 0
\end{array}\right)
\qquad
W = \left(\begin{array}{ccc}
-i |u|^2 & \bar u_x\\
-u_x & i |u|^2
\end{array}\right) = \sigma_3 (U_x - i U^2).
\enq

It is important to note, that the zero-curvature condition for \eqref{NLS} can also be rewritten as a conjugate system for a {\em different} $2\times 2$ matrix-valued function $\Phi = \Phi(x,t)$ and different spectral parameter $\mu$ (independent of $\l$):
\beq \label{Lax_b}
\begin{split}
\Phi_x & = i M  \Phi \sigma_3 - i \Phi U , \\
\Phi_t & = - 2 i M^2 \Phi \sigma_3 + 2 i M \Phi U - \Phi W.
\end{split}
\enq

Furthermore, it is easy to see that the matrix $U$ can be rewritten as
\beqn
U = \Re(u) \left(\begin{array}{ccc}
0 & 1\\
1 & 0
\end{array}\right) +  \Im (u) \left(\begin{array}{ccc}
0 & -i\\
i & 0
\end{array}\right) = \Re (u) \sigma_1 + \Im (u) \sigma_2,
\enqn
where $\sigma_1$ and $\sigma_2$ are the first and second Pauli matrices, and just like $\sigma_3$, they are Hermitian and unitary. Thus, $U$ is itself Hermitian. Similarly, it is easy to show that the matrix $W$ is skew-Hermitian, i.e.:
\beq \label{Hermitian}
U^+ = U, \qquad W^+ = -W,
\enq
where $~^+$ indicates the conjugate transpose. \eqref{Hermitian} implies that for known $\Psi$ and $\Lambda$ the choice
\beq \label{Phi_Psi}
\Phi = \Psi^+, \qquad M = - \Lambda^+,
\enq
will automatically satisfy the conjugate system \eqref{Lax_b}.

Next, we will need an additional tool, designed with the Darboux transformation in mind: a closed 1-form $\Omega=\Omega(\Phi,\Psi)$, that satisfies both the condition
\beq \label{Omega}
\Phi \Psi = i (M \Omega + \Omega \Lambda),
\enq
and the differential equation
\beq\label{Omega_x}
\Omega_x = \Phi \sigma_3 \Psi.
\enq
In order to close the 1-form, that is for it to satisfy the compatibility condition
\beq \label{Omega_compat}
\Omega_{tx}=\Omega_{xt},
\enq
the function $\Omega$ should satisfy the following easily verifiable condition:
\beq \label{Omega_t}
\Omega_t = 2 i (\Phi_x \Psi -\Phi \Psi_x) - 2 \Phi U \Psi = 2(\Phi \sigma_3 \Psi \Lambda - M \Phi \sigma_3 \Psi) + 2 \Phi U \Psi.
\enq

Moving ever closer to the task of construction of the Darboux transformation, let us assume that in addition to $\Psi$ and $\Phi$, we also have two other functions $\Psi_1$ and $\Phi_1$, that are also solutions to \eqref{Lax_a}, \eqref{Lax_b} albeit for the different spectral matrices $\Lambda_1$ and $M_1$ correspondingly. These functions produce two supplementary matrices $\tau$ and $\sigma$:
\beq \label{tau_sigma}
\tau = \Psi_1 \Lambda_1 \Psi_1^{-1}, \qquad \sigma = \Phi_1^{-1} M_1 \Phi_1,
\enq
that are notable for satisfying the following conditions (with brackets denoting the commutator):
\beq \label{tau_system}
\begin{split}
\tau_x &= i [\sigma_3, \tau] \tau + i [U, \tau], \\
\tau_t &= 2 i [\sigma_3,\tau] \tau^2 + 2 i [U,\tau] \tau +[W, \tau],
\end{split}
\enq
and
\beq \label{sigma_system}
\begin{split}
\sigma_x &= i \sigma [\sigma, \sigma_3] + i [U, \sigma], \\
\sigma_t &= 2 i \sigma^2 [\sigma_3,\sigma] - 2 i \sigma [U,\sigma] + [W, \sigma].
\end{split}
\enq

With all these pieces now in place we are finally free to define a new Darboux transformation, that would utilize both the supplementary one-form $\Omega$ and the support function $\Psi_1$:
\beq \label{Darboux_plus1}
\begin{split}
\Phi \to \Phi^{(+1)} &= \Omega(\Phi, \Psi_1) \Psi_1^{-1} \\
\Phi_1 \to \Phi^{(+1)} &= \Omega(\Phi_1, \Psi_1) \Psi_1^{-1}.
\end{split}
\enq
Such transformation will naturally affect all the remaining ingredients of \eqref{Lax_a} and \eqref{Lax_b}, transforming them via the following easily verifiable formulas:
\beq \label{Darboux_plus2}
\begin{split}
\Psi \to \Psi^{(+1)} &= \Psi \Lambda - \tau \Psi \\
U \to U^{(+1)} &= U + [\sigma_3, \tau] = U +2\sigma_3 \tau \\
W \to W^{(+1)} &= W + 2i \left(U^{(+1)} \tau - \tau U\right) = \sigma_3 \left(U^{(+1)}_x - i (U^{(+1)})^2\right).
\end{split}
\enq

We would like to point out at this step that the Darboux transformation \eqref{Darboux_plus1}, \eqref{Darboux_plus2} all rely on the support function $\Psi_1$, which was a particular solution of \eqref{Lax_a} with the spectral matrix $\Lambda=\Lambda_1$. This, however, is but one of the possibilities. The alternative way to define our transformation would be via the function $\Phi_1$ from \eqref{Lax_b}, thus producing the following system:
\beq \label{Darboux_minus}
\begin{split}
\Psi \to \Psi^{(-1)} &= \Phi_1^{-1} \Omega(\Phi_1, \Psi) \\
\Psi_1 \to \Psi^{(-1)} &= \Phi_1^{-1} \Omega(\Phi_1, \Psi_1) \\
\Phi \to \Phi^{(-1)} &= M \Phi  - \Phi \sigma \\
U \to U^{(-1)} &= U + [\sigma_3, \sigma].
\end{split}
\enq

It is important to point out here that in order for the formulas for $\Phi^{(+1)}$ and $\Phi_1^{(+1)}$ from \eqref{Darboux_plus1}, as well as $\Psi^{(-1)}$ and $\Psi_1^{(-1)}$ from \eqref{Darboux_minus} to be satisfied, the condition \eqref{Omega} should hold. In particular, for $\Phi^{(+1)}$ to be true, one should have
\beq \label{Phi_Psi1}
\Phi \Psi_1 = i \left(M \Omega(\Phi, \Psi_1) + \Omega(\Phi, \Psi_1) \Lambda_1\right),
\enq
whereas the prerequisite for the Darboux transformation $\Psi_1 \to \Psi_1^{(-1)}$ is
\beq \label{Phi1_Psi1}
\Phi_1 \Psi_1 = i \left(M_1 \Omega(\Phi_1, \Psi_1) + \Omega(\Phi_1, \Psi_1) \Lambda_1\right).
\enq

Suppose the \eqref{Phi1_Psi1} indeed holds. Then it is possible to utilize both positive and negative Darboux transforms and introduce a new {\em binary Darboux transformation} (binary DT), which can be defined in two ways:
\beq \label{binaryDT}
\begin{split}
U &\to U^{(+1)} \to U^{(+1,-1)} \\
U & \to U^{(-1)} \to U^{(-1,+1)},
\end{split}
\enq
where the first of binary DT is defined as:
\beq \label{U_plus-minus}
\begin{split}
U^{(+1,-1)} &= U^{(+1)} + \left[\sigma_3, \sigma^{(+1)}\right], \\
\sigma^{(+1)} &= \left(\Phi_1^{(+1)}\right)^{-1} M_1 \Phi_1^{(+1)} = \Psi_1 \Omega^{-1} (\Phi_1, \Psi_1) M_1 \Omega(\Phi_1, \Psi_1) \Psi_1^{-1},
\end{split}
\enq
which can alternatively be rewritten as
\beq \label{U_plus-minus_simpl}
U^{(+1,-1)}=U+\left[\sigma_3, \tau+\sigma^{(+1)}\right] = U - i \left[\sigma_3, G_{11}\right],
\enq
where for simplicity we have introduced a new matrix-valued function $G_{11}$ defined as
\beq \label{G11}
G_{11} = \Psi_1 \Omega^{-1}(\Phi_1, \Psi_1) \Phi_1.
\enq

Most astonishingly, by repeating the exact same calculations for the second binary DT $U^{(-1,+1)}$ will produce exactly the same result, thus producing the following extremely important result:
\beq \label{U_is_U}
U^{(+1,-1)} = U^{(-1,+1)}.
\enq
This means that the $+1$ and $-1$ DT's actually {\em commute} with each other and their order is inessential. We can therefore introduce the following notation:
\beq \label{U_n-m}
U \to U^{(+n,-m)}, \qquad n, m \in \N,
\enq
where $n$ and $m$ denote the amount of ``positive'' and ``negative'' Darboux transformations applied to $U$.\footnote{Naturally, this notation allows to define $U$ itself as $U^{(+0,-0)}$ and the individual ``positive and ``negative'' DT's as $U^{(+1,-0)}$ and $U^{(+0,-1)}$ correspondingly.}


\section{The Reduction Restriction (and a First Snag)} \label{sec:Reduction_Restriction}

In the previous chapter we have demonstrated the existence of not just one but {\em two} Darboux transformations: $U \to U^{(+1)}$ and $U \to U^{(-1)}$ and have shown that together they form a binary DT $U \to U^{(+1,-1)}$, and, indeed, a whole slew of iterative binary DT's that we have denoted as $U^{(+n,-m)}$. However, the corresponding DT's all share a similar problem: in general they do not respect the Hermitian condition \eqref{Hermitian}. In this chapter we would like to amend this little snag by restricting our attention to just those binary DT's that produce the Hermitian matrices, that is, satisfy the condition
\beq \label{reduction_restriction}
\left(U^{(+n,-m)}\right)^+= U^{(+n,-m)}, \qquad \text{for} \ \forall ~n,m \in \N,
\enq
which we will henceforth call the {\em reduction restriction} condition.

We will begin by examining the DT we introduced first: the $U^{(+1)}$. To be specific, let us define the individual elements of matrices $\Psi_1$ and $\Lambda_1$ as:
\beq \label{Psi1_Lambda1}
\Psi_1 = \left(\begin{array}{cc}
\psi_1 & \psi_2\\
\phi_1 & \phi_2
\end{array}\right), \qquad \Lambda_1 = \left(\begin{array}{ccc}
\mu & 0\\
0 & \bar \mu
\end{array}\right).
\enq

Fortunately, we don't have to work with all four individual elements of matrix $\Psi_1$, since the exist the following famous (and easily verifiable) reduction \cite{Salle}:
\beq \label{psi_reduced}
\psi_1 = \bar \phi_2, \qquad \psi_2 = - \bar \phi_1.
\enq
Moreover, we can significantly simplify our calculations by choosing as simple seed solution $u(t,x)$ of \eqref{NLS} as possible. In our case we will work with the periodic solution
\beq \label{u_a_A}
u = A \eb^{i S}, \qquad S = a x + (2A^2-a^2) t, \qquad a, A \in \R.
\enq

From \eqref{Psi1_Lambda1}, \eqref{psi_reduced} and \eqref{u_a_A} substituted into \eqref{Lax_a} we immediately obtain the following partial differential equations on $\phi_1$ and $\phi_2$:
\beq \label{phi_system}
\begin{split}
{\phi_1}_x & = - i \mu \phi_1 + i u \bar \phi_2 \\
{\phi_2}_x & = -i \bar \mu \phi_2 - i u \bar \phi_1 \\
{\phi_1}_t & = i \left(A^2 - 2 \mu^2 \right) \phi_1 - i u \left(2\mu - a \right) \bar \phi_2 \\
{\phi_2}_t &= i \left(A^2 - 2 \bar\mu^2 \right)\phi_2 + i u \left(a - 2\bar \mu\right) \bar \phi_1.
\end{split}
\enq

Applying the DT \eqref{Darboux_plus2} will produce a new NLS solution $u^{(1)}$ that will have a form
\beq \label{u1}
u^{(1)} = u + \frac{2(\bar \mu - \mu) \phi_1 \phi_2}{|\phi_1|^2 + |\phi|^2},
\enq
and a similar formula for $\bar u^{(1)}$, indeed producing the matrix $U^{(+1)}$ in a Hermitian form, and therefore substantiating the previous reduction \eqref{psi_reduced}. The particular solutions for $\phi_1$ and $\phi_2$ for the resulting problem will be of the form \cite{Salle}
\beq \label{f_i}
\phi_i = f_i \eb^{i S/2}, \qquad i=1,2,
\enq
where $f_i$ are real-valued and $S$ is the same as in \eqref{u_a_A}. After the substitution of \eqref{f_i} into the linear system \eqref{phi_system} we will have a number of possibilities open; one that we are particularly interested in corresponds to the case when the roots of the characteristic equations on $f_i$ are both equal to zero, which happens when
\beq \label{mu_for_Peregrin}
\mu = -\frac{a}{2} \pm i A.
\enq
Not surprisingly, the resulting functions $f_i$ will be linear in terms of both $x$ and $t$ variables:
\beq\label{ffi}
f_1=bx+2b(i A-a)t+c,\qquad f_2=2{\bar b}(i a-A)t-i{\bar b} x+i\left(\frac{{\bar b}}{A}-{\bar c}\right),
\enq
What {\em is} interesting, is that upon the Darboux transformation \eqref{u1} and the subsequent simplifications, we will end up with a {\em localized} solution: the Peregrine breather, which is essentially a Peregrine soliton on a the background of a planar wave \eqref{u_a_A},
\beq\label{Pere1}
\begin{split}
u^{(1)} & = A{\rm e}^{iS}\left(-1+\frac{2\left(1+4i t A^2\right)}{2A^2\eta^2-2A\eta+8A^4t^2+1}\right),\\
\\
|u^{(1)}|^2 & =  A^2+\frac{8A^3\left(4A^3t^2+\eta-A\eta^2\right)}{\left(2A^2\eta^2-2A\eta+8A^4t^2+1\right)^2},
\end{split}
\enq
with $\eta=x-2at$, $b=1$, $c=0$ and we choose the down sign (''-'') in \eqref{mu_for_Peregrin}. Thus, we actually end with a very simple and straightforward mechanism for ``construction'' of the Peregrine solitons for NLS. Reverting the steps taken in Section \eqref{sec:Intro} and returning from $u$ back to magnetization $\vec M$, will therefore yield us what the authors of \cite{ZLLWFL} has called the magnetic rogue wave (however, as we can now see, theirs was basically just a rediscovery of a well-known result of Matveev and Salle \cite{Salle}, merely redressed for the LLG).

Unfortunately, this simple algorithm is literally a single-shot weapon! A stumbling block here is the condition \eqref{mu_for_Peregrin}. Although it allows us to produce a Peregrine breather after one DT, for our purposes it is not enough. In order to build a multisoliton Peregrine-like solution, we should use multiple iterations of DT, and so we must have at least {\em two} linearly independent Peregrine solitons. But each one of them would have to be constructed with a spectral restriction \eqref{mu_for_Peregrin} in mind; and it can be shown that, regardless of the sign in \eqref{mu_for_Peregrin}, the resulting Peregrine solutions will always be linearly dependent, thus producing nothing but zero after the second iteration of DT.

One way around this obstacle has been proposed by Dubard and Matveev in \cite{Dubard_Matveev} where they have used the relationship existing between the focusing NLS and the Kadomtsev-Petviashvili I (KP-I) equation to produce a multi-rogue waves solutions for NLS (and, subsequently, a family of localized rational solutions of KP-I). Hover, in this article we will adopt a different strategy and demonstrate that it is actually possible to develop a strategy for construction of a multi-rogue wave profile NLS solution while remaining entirely in the framework of Darboux transformation for NLS. The key here is to use not a DT, but a {\em binary DT}!


\section{Understanding the Binary DT: from the Stationary Schr\"odinger Equation to KdV} \label{sec:Binary_DT}

Before we proceed it would be beneficial to take a glance at a theory of binary DT for a simpler equation than NLS: the stationary Schr\"odinger equation. Suppose $\psi$ and $\phi$ are some linearly independent solutions for the Schr\"odinger equation with the same potential $v = v(x)$, albeit with different spectral parameters $\lambda$ and $\mu$:
\beq \label{Schrodinger}
\begin{split}
\psi_{xx} &= (v-\l) \psi \\
\phi_{xx} &= (v-\mu) \phi.
\end{split}
\enq
Then out of these two solutions one can construct a new solution $\psi^{(1)}$ to the Schr\"odinger equation
\beqn
\psi^{(1)}_{xx} = (v^{(1)} - \l) \psi^{(1)},
\enqn
where the new solution and a new potential $v^{(1)}$ are calculated via the Darboux transformation:
\beq \label{Darboux_Schrodinger}
\psi \to \psi^{(1)} = \psi_x -\frac{\phi_x}{\phi} \psi, \qquad v \to v^{(1)} = v - 2\left(\ln \phi\right)_{xx}.
\enq

It is quite apparent from \eqref{Darboux_Schrodinger} that the linear independence of $\phi$ and $\psi$ is a crucial condition, as its violation produces only a trivial solution $\psi^{(1)} \equiv 0$. This condition is automatically satisfied when we choose $\l \neq \mu$. But what if they are identical, as was the case with the Peregrine solitons in the previous chapter?.. Is DT useless in this case?.. Actually, the answer is {\em no}, and it has to do with the fact that the Schr\"odinger equation is a second order linear O.D.E., and therefore must necessary have two linearly independent solutions for any value of a spectral parameter. \footnote{Although only one of them can belong to $L_2$ space.} Suppose, for example, that we are looking for a solution $\tilde \phi$ that is linearly independent of $\phi$ but satisfies the same equation with the same spectral parameter. Using the well-known identity
\beqn
\frac{d}{dx}\left(\frac{\tilde \phi}{\phi}\right) = \frac{\Delta(\tilde \phi, \phi)}{\phi^2},
\enqn
where $\Delta$ is a Wronskian of the solutions $\tilde \phi$ and $\phi$, and the fact that for the Schr\"odinger equation the Wronskian of two solutions is constant\footnote{This is the result of a following easily verifiable fact: for any homogeneous linear O.D.E.
\beqn
y'' + p(x) y' + q(x)y = 0,
\enqn
the Wronskian of two solutions $y_1$ and $y_2$ will always satisfy the condition $\ln \Delta(y_1,y_2) = \int p(x) dx$.} immediately means that
\beq \label{phi_tilde}
\tilde \phi = \phi \int \frac{dx}{\phi^2}.
\enq
This new solution has a distinction of being linearly independent of the original $\phi$. Therefore, we can apply the Darboux transformation \eqref{Darboux_Schrodinger} to $\phi$ and $\tilde \phi$ and in doing so produce a nontrivial function that we will call $\tilde \phi^{(1)}$:
\beq
\tilde \phi^{(1)} = \tilde \phi_x - \frac{\phi_x}{\phi} \tilde \phi = \frac{1}{\phi}.
\enq
As we have discussed above, this function will be a solution to a Schr\"odinger equation with a new potential $u^{(1)}$. This new equation will also have a {\em second} solution, linearly independent of $\tilde \phi^{(1)}$, that can be calculated using \eqref{phi_tilde}. This new solution that we will call $\phi^{(1)}$ will have a form:
\beq \label{Schrodinger_phi1}
\phi^{(1)} = \frac{1}{\phi} \int \phi^2 dx.
\enq
Thus, we end up with two linearly independent solutions of a new Schr\"odinger equation with a new potential $v^{(1)}$. But then we can use them in the Darboux transformation once again, and produce another solution for yet another new potential $v^{(2)}$:
\beq \label{binary-0}
v^{(2)} = v^{(1)}-2\left(\ln \phi^{(1)}\right)_{xx} = v - 2\frac{d^2}{dx^2} \ln \left(\int \phi^2 dx\right),
\enq
and it is this transformation that is called the {\em binary Darboux transformation}.

It is interesting to note that \eqref{binary-0} allows one to construct positons and negatons solutions of the KdV equation directly. Such solutions were obtained in a set of articles  (see for example \cite{Matveev1}, \cite{Matveev2}) via some generalization of DT, namely via the Darboux transformations complemented by the differentiation with respect to a spectral parameter. Importantly, it was the similar approach used in \cite{Dubard_Matveev} that eventually led to construction of new multi-rogue waves. On  the other hand, as we shall soon see, both positon (negaton) solutions of the KdV equations and the multi-rogue waves solutions of the focusing NLS equations might all be obtained just by the binary DT without any additional differentiation. In the remainder of this section, we'll concentrate on showing this for the KdV, so that when we return back to the discussion of NLS equations in next sections we will already have a good point of reference.

The KdV equation ($v=v(x,t)$) has the well known form:
\beqn
v_t-6vv_x+v_{xxx}=0,
\enqn
and may be obtained from its Lax pair, which consists of one of the equations from the system \eqref{Schrodinger} (let it be the equation for the $\phi=\phi(x,t)$) and an additional evolutionary equation:
\beq\label{phit}
\phi_t=2\left(v+2\mu\right)\phi_x-v_x\phi.
\enq
If we put $v=0$, $\mu=-\kappa^2$ and $\phi=\cosh\eta$ with $\eta=\kappa(x-4\kappa^2t)$
then \eqref{Darboux_Schrodinger} results in the famous one-soliton solution $v^{(1)}=-2\kappa^2\sec^2\eta$. The dressed solution of Lax pair with the same value of the spectral parameter may be calculated with the help of \eqref{Schrodinger_phi1} with one caveat -- one has to correctly define the limits of integration in order to be sure that $\phi^{(1)}(x,t)$ is indeed a solution of \eqref{phit} with $v\to v^{(1)}$. We choose the upper limit as $x$ and a lower limit as $\omega(t)$ where the function $\omega(t)$ must be obtained. So:
\beq\label{phi11}
\phi^{(1)}=\frac{1}{\phi} \int_{\omega(t)}^x \phi^2 dx=\frac{1}{\kappa \cosh \eta}\int_{\eta_{_1}(t)}^{\eta} \cosh^2\eta' d\eta'=\frac{2\eta+\sinh 2\eta-2\eta_{_1}(t)-\sinh 2\eta_{_1}(t)}{4\kappa \cosh \eta},
\enq
with $\eta_{_1}(t)=\kappa\left(\omega(t)-4\kappa^2t\right)$. Substituting \eqref{phi11} into the \eqref{phit} one gets the simple first order differential equation  for the unknown $\eta_{_1}(t)$ which may be integrated, so (the integration constant being omitted for the sake of brevity)
\beq\label{eta1}
2\eta_{_1}(t)+\sinh 2\eta_{_1}(t)=16 \kappa^3 t.
\enq
Substituting \eqref{eta1} into the \eqref{phi11} and then using \eqref{binary-0} (with the correct limits of integration) we get exactly one negaton solution from the \cite{Matveev2} (we again omit the constants of integration):
\beq\label{negaton}
v^{(2)}=-2\frac{\partial^2}{\partial x^2}\ln\left(\sinh 2\eta+2\kappa\beta\right),
\enq
with $\beta=x-12 \kappa^2 t$. To obtain the positon solution one should simply repeat the described procedure for a positive value of $\mu$.


\section{Binary DT and NLS} \label{sec:Binary_DT_NLS}

Let's say we are planning to extend the approach discussed in the previous chapter to the task of construction of a binary DT for NLS.  We will begin by imposing the already familiar restrictions (recall that $^+$ denotes the conjugate transpose):
\beqn \label{Phi_Psi1}
\Phi = \Psi^+, \qquad M = - \Lambda^+,
\enqn
and observing that the closed form $\Omega$ should then satisfy the following condition:
\beq \label{psi+psi}
\Psi^+ \Psi = i (\Omega \Lambda -\Lambda^+ \Omega),
\enq
where, just as before, we will choose
\beq \label{Psi+Lambda}
\Psi = \left(\begin{array}{cc}
\bar \phi_2 & - \bar \phi_1\\
\phi_1 & \phi_2
\end{array}\right), \qquad \Lambda_1 = \left(\begin{array}{cc}
\mu & 0\\
0 & \bar \mu
\end{array}\right),
\enq
and we denote the entries of the matrix $\Omega$ as:
\beq \label{Omega_matrix}
\Omega=\Psi_1 = \left(\begin{array}{cc}
\Omega_{11} & \Omega_{12}\\
\Omega_{21} & \Omega_{22}
\end{array}\right).
\enq
Taken together, \eqref{Omega_matrix}, \eqref{Psi+Lambda} and the condition \eqref{Phi_Psi1} immediately provides the diagonal values of $\Omega$:
\beq \label{Omega_diagonal}
\Omega_{11} = -\Omega_{22} = \frac{i}{\bar \mu - \mu} \left(|\phi_1|^2 + |\phi_2|^2\right).
\enq

The non-diagonal elements are a bit trickier to obtain; finding them requires solving the differential equations \eqref{Omega_t} and \eqref{Omega_x}, that lead to the following system:
\beq
\begin{split}
\Omega_{12, x} &= -2\phi_2 \bar \phi_1 \\
\Omega_{21, x} &= -2 \bar \phi_2 \phi_1 \\
\Omega_{12, t} &= -8 \bar \mu \phi_2 \bar \phi_1 +2 \bar u \phi_2^2 - 2u\bar \phi_1^2 \\
\Omega_{21, t} &= -8 \mu \phi_1 \bar \phi_2 + 2 u \bar \phi_2^2 - 2 \bar u \phi_1^2,
\end{split}
\enq
which can be integrated for any particular $\phi_1$ and $\phi_2$. In particular, evaluating the matrix $\Omega$ for the plane wave \eqref{u_a_A} with the functions $\phi_i$ defined as in \eqref{f_i} and \eqref{ffi} one gets:
\beq\label{Omsol}
\begin{split}
\Omega_{11} & = -\frac{\eta^2}{A}+\frac{\eta}{A^2}-4At^2-\frac{1}{2A^3},\\
\Omega_{12} & =c_{12}+i\left(\frac{2\eta}{3}-\frac{1}{A}\right)\eta^2-\frac{16 A^3 t^3}{3}-4i A\left(2A\eta-1\right)t^2+2\left(2A\eta^2-2\eta-\frac{1}{A}\right)t,\\
\Omega_{21} & = {\bar \Omega}_{12},\qquad \Omega_{22}=-\Omega_{11}.
\end{split}
\enq
and performing the binary DT, described in Section \ref{sec:Lax}, will produce the $n=2$ multi-rogue wave solutions of \cite{Dubard_Matveev}. However, the plane wave is just one of possible seed solutions; what would happen should we choose, for example, a {\em zero} solution $u=0$?.. Would this produce the same already known multi-soliton rogue waves?.. As we shall see, the answer is no; we will instead get something rather unexpected: a rogue wave that arises during the impact of two positon solitons!

\section{The Positon-produced Rogue Wave} \label{sec:Positons}

In this Section we’ll take a particular look at what happens when we use our approach on the one-soliton background instead of a plane wave. As we shall see, in this case the binary DT produces a new rogue wave-like solutions of the NLS, whose scattering profile consists of two ``positon'' waves.

The first step is traditional: to construct a one soliton solution we start at a zero background $u=0$. In this case the LA-pairs equations have an extremely simple form
\begin{equation} \label{Pos-1}
\phi_{1,x} = -i \mu \phi_1, \qquad \phi_{2,x} = -i \bar \mu \phi_2, \qquad \phi_{1,t}=-2i\mu^2\phi_1, \qquad \phi_{2,t}=-2i\left(\bar \mu \right)^2\phi_2.
\end{equation}
Solving \eqref{Pos-1} yields
\begin{equation} \label{Pos-2}
\phi_1=C_1{\rm e}^{-i\left(\alpha x+2(\alpha^2-\beta^2)t\right)+\beta\xi},\qquad
\phi_2=C_2{\rm e}^{-i\left(\alpha x+2(\alpha^2-\beta^2)t\right)-\beta\xi},
\end{equation}
with
$$
\mu=\alpha+i\beta,\,\, \bar \mu = \alpha-i\beta,\,\,\xi=x+4\alpha t, \ \ C_1,C_2 \in \R.
$$
So, after one DT we end up with the following one soliton solution:
\begin{equation} \label{Pos-3}
u^{(+1)}=-\frac{2i\beta C_1C_2}{|C_1||C_2|}\frac{{\rm e}^{-2i\left(\alpha x+2\left(\alpha^2-\beta^2\right)t\right)}}{\cosh\left(2\beta(x+4\alpha t)+\log\frac{|C_1|}{|C_2|}\right)}.
\end{equation}
As we have discussed in Section \ref{sec:Lax}, in order to dress \eqref{Pos-3} via the binary DT one should start by finding the matrix $\Omega_{ik}$. The calculations results in
\begin{equation} \label{Pos-4}
\Omega_{12}=c-2x-8\bar \mu t= \bar \Omega_{21},\qquad \Omega_{22}=-\Omega_{11}=\frac{1}{\beta}\cosh(2\beta\xi),
\end{equation}
So
\begin{equation} \label{Pos-sol}
u^{(+1,-1)}=\frac{4{\rm e}^{-2i \theta} \left(8\beta t \cosh(2 \beta \xi) + i \left(2\xi\sinh(2\beta\xi)-\frac{\cosh(2\beta\xi)}{\beta}\right)\right)}{\left(\frac{1}{\beta} \cosh(2\beta\xi)\right)^2+4\left(\xi^2+16\beta^2 t^2\right)},
\end{equation}
with $\theta=\alpha x+2\left(\alpha^2-\beta^2\right)t$. It is immediately obvious from \eqref{Pos-sol} that this function is even with respect to the variable $\xi$:
\begin{equation}
u^{(+1,-1)}(-\xi,t)= u^{(+1,-1)}(\xi,t);
\label{Pos-chyot1}
\end{equation}
the importance of this seemingly innocent fact will become apparent a bit later on.

It would be beneficial now to switch from $(\xi,t)$ to a new coordinate system $(x,t)$, that moves with a velocity $v=-4\alpha$ relative to the old one. In this new system the absolute value of $u^{(+1,-1)}$ turns into
\begin{equation}
\left|u^{(+1,-1)}\right|^2=16~\frac{64\beta^2t^2\cosh^2(2\beta x)+\left(2x\sinh(2\beta x)-\frac{1}{\beta}\cosh(2\beta x)\right)^2}{\left(\frac{1}{\beta}\cosh(2\beta x)\right)^2+4\left(x^2+16\beta^2 t^2\right)},
\label{Pos-sq}
\end{equation}
which is a function that is even with respect to not just one but {\em both} variables!
\begin{equation}
\left|u^{(+1,-1)}(x,t)\right|^2 = \left|u^{(+1,-1)}(x,t)\right|^2 = \left|u^{(+1,-1)}(-x,-t)\right|^2 = \left|u^{(+1,-1)}(x,-t)\right|^2.
\label{Pos-chyot2}
\end{equation}

\begin{figure}
\begin{center}
\includegraphics[width=0.6\columnwidth]{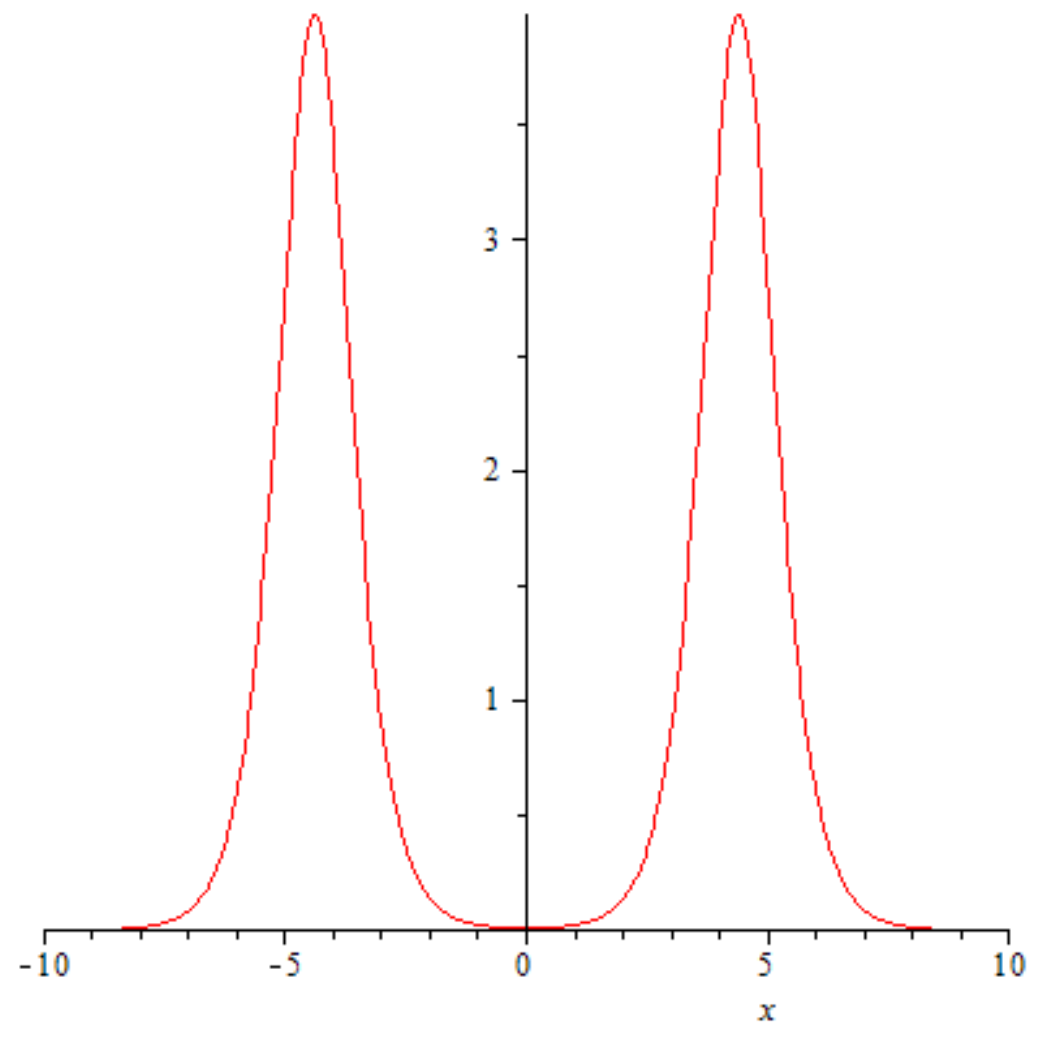}
\caption{\label{fig1} A graph of function $\left|u^{(+1,-1)}(x,t)\right|^2$ at $t=-5$. The parameter $\beta=1$.}
\end{center}
\end{figure}

The dynamics of the solution \eqref{Pos-sq} is this: for large absolute values of the variable $t$ it describes two symmetric ``solitons'' that are slowly approaching each other (see Fig \ref{fig1}).

\begin{figure}
\begin{center}
\includegraphics[width=0.6\columnwidth]{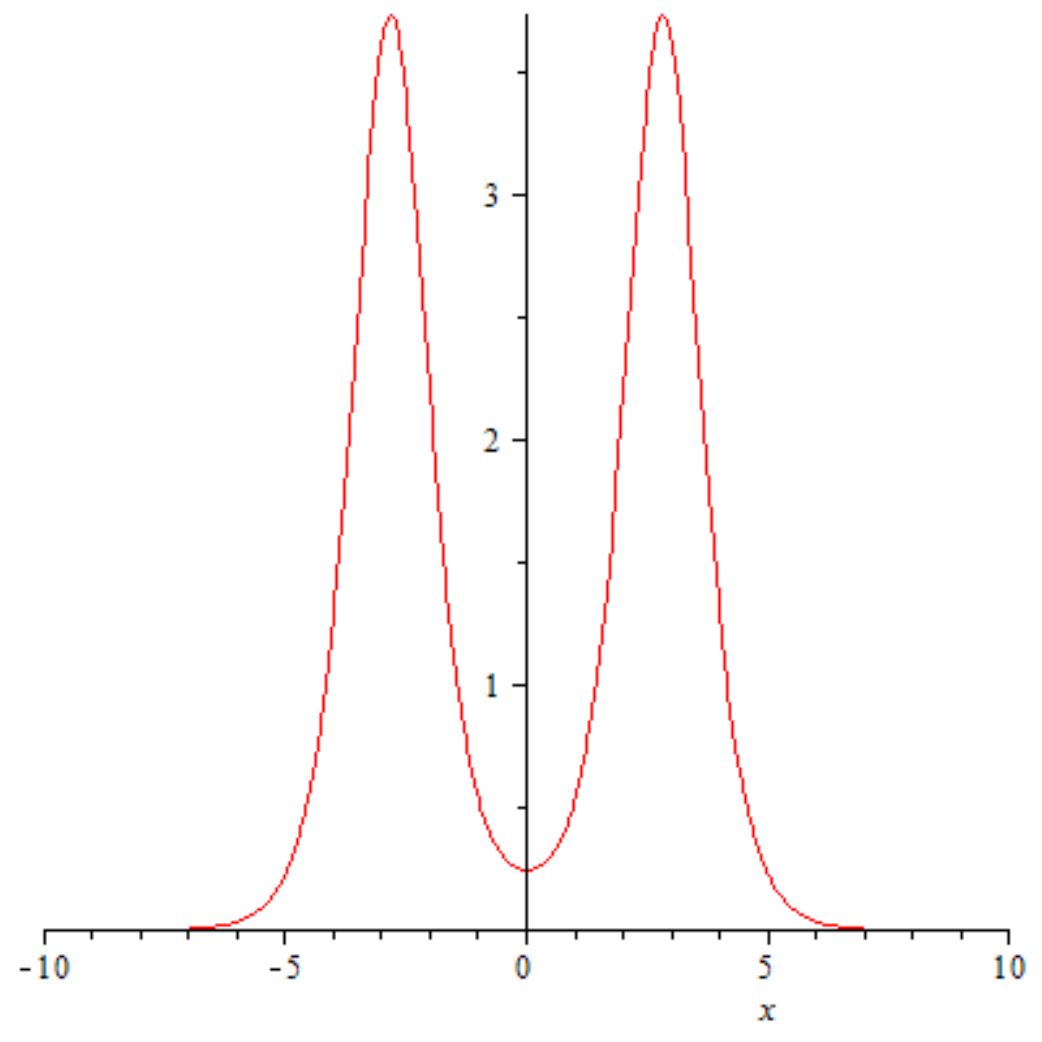}
\caption{\label{fig2} A graph of function $\left|u^{(+1,-1)}(x,t)\right|^2$ at $t=-1$.}
\end{center}
\end{figure}

\begin{figure}
\begin{center}
\includegraphics[width=0.6\columnwidth]{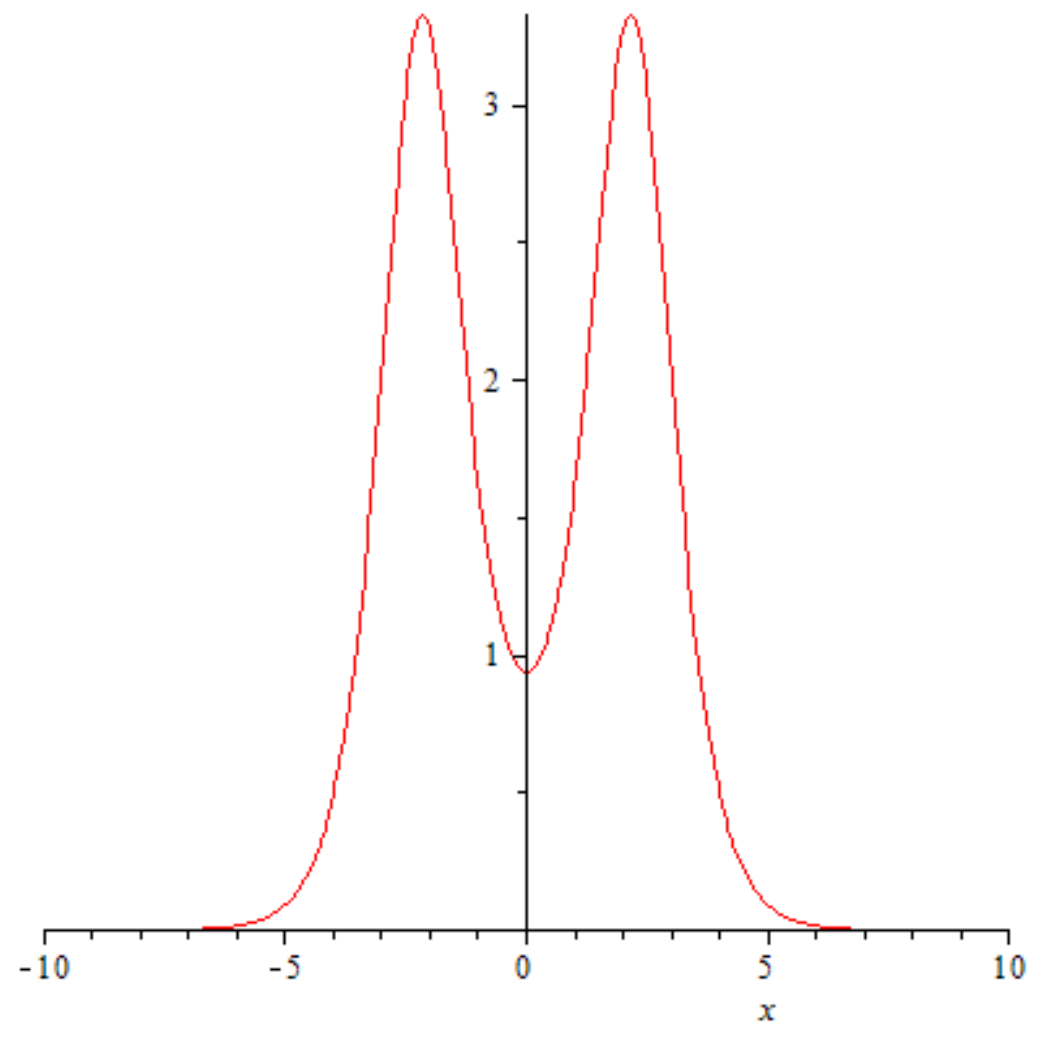}
\caption{\label{fig3} A graph of function $\left|u^{(+1,-1)}(x,t)\right|^2$ at $t=-0.5$.}
\end{center}
\end{figure}

\begin{figure}
\begin{center}
\includegraphics[width=0.6\columnwidth]{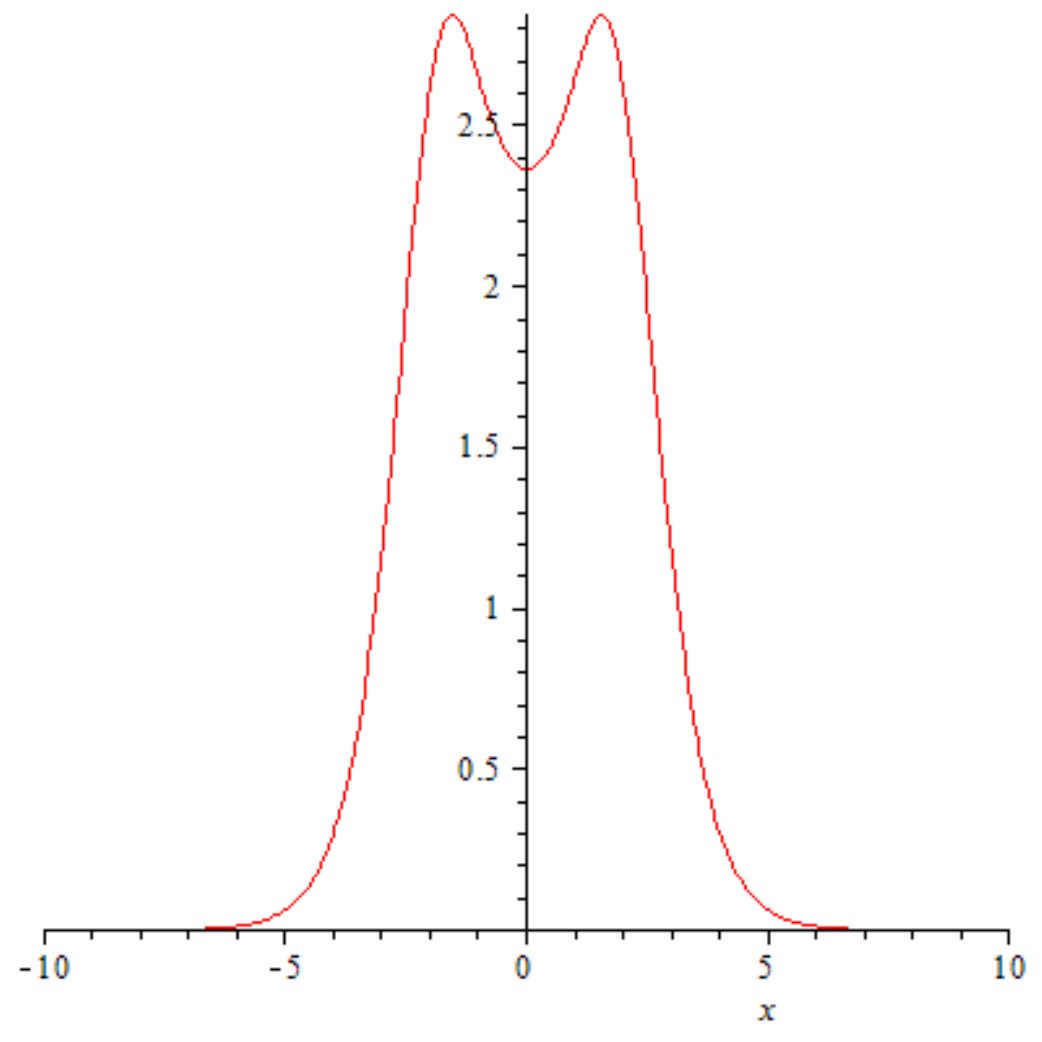}
\caption{\label{fig4} A graph of function $\left|u^{(+1,-1)}(x,t)\right|^2$ at $t=-0.3$.}
\end{center}
\end{figure}

\begin{figure}
\begin{center}
\includegraphics[width=0.6\columnwidth]{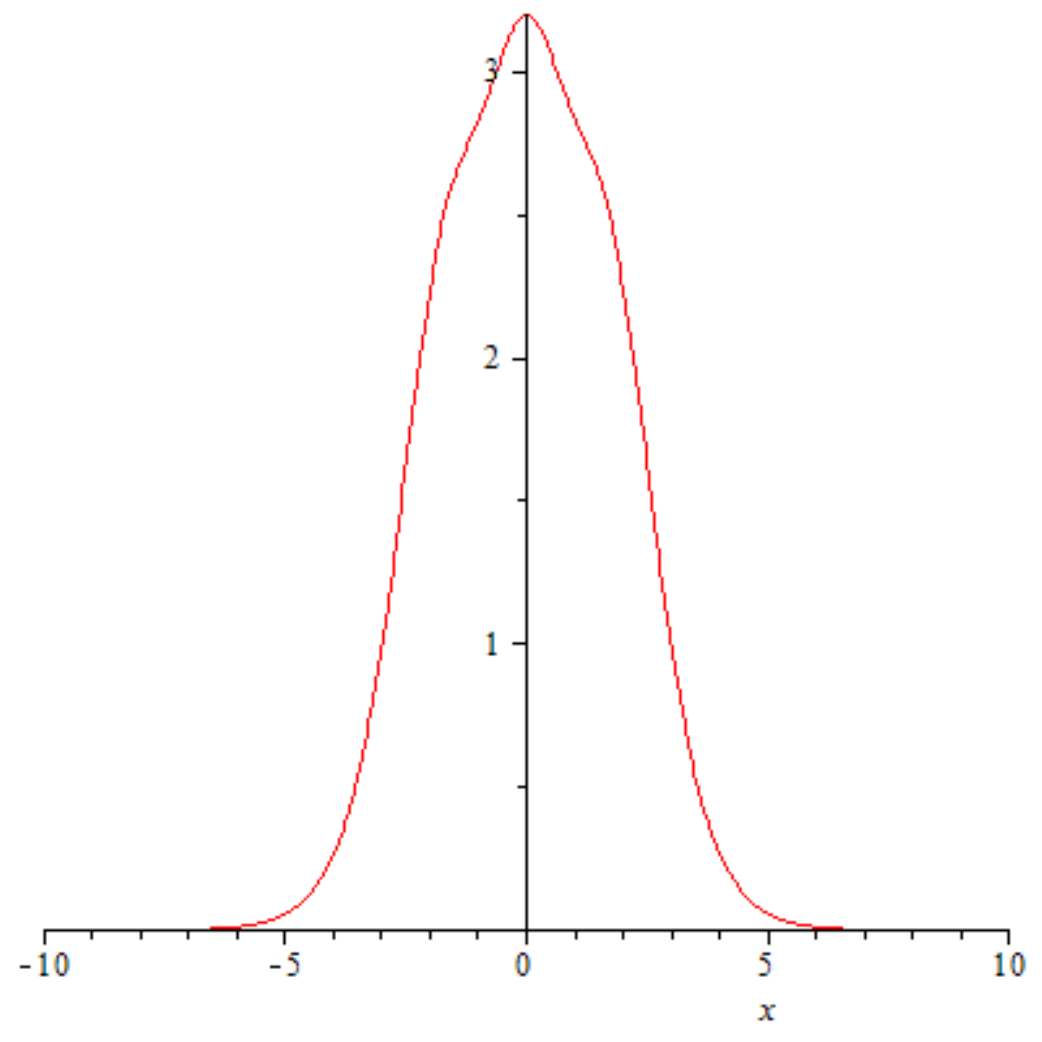}
\caption{\label{fig5} A graph of function $\left|u^{(+1,-1)}(x,t)\right|^2$ at $t=-0.25$.}
\end{center}
\end{figure}

They impact each other at $t=0$ and $x=0$, and it happens in a very unusual fashion: the amplitudes of the solitons crashing into each begin a rapid decrease (Figs \ref{fig2}-\ref{fig4}), at the very point of impact ($x=0$ and its immediate neighbourhood) a new ephemeral peak emerges (Fig \ref{fig5}).

\begin{figure}
\begin{center}
\includegraphics[width=0.6\columnwidth]{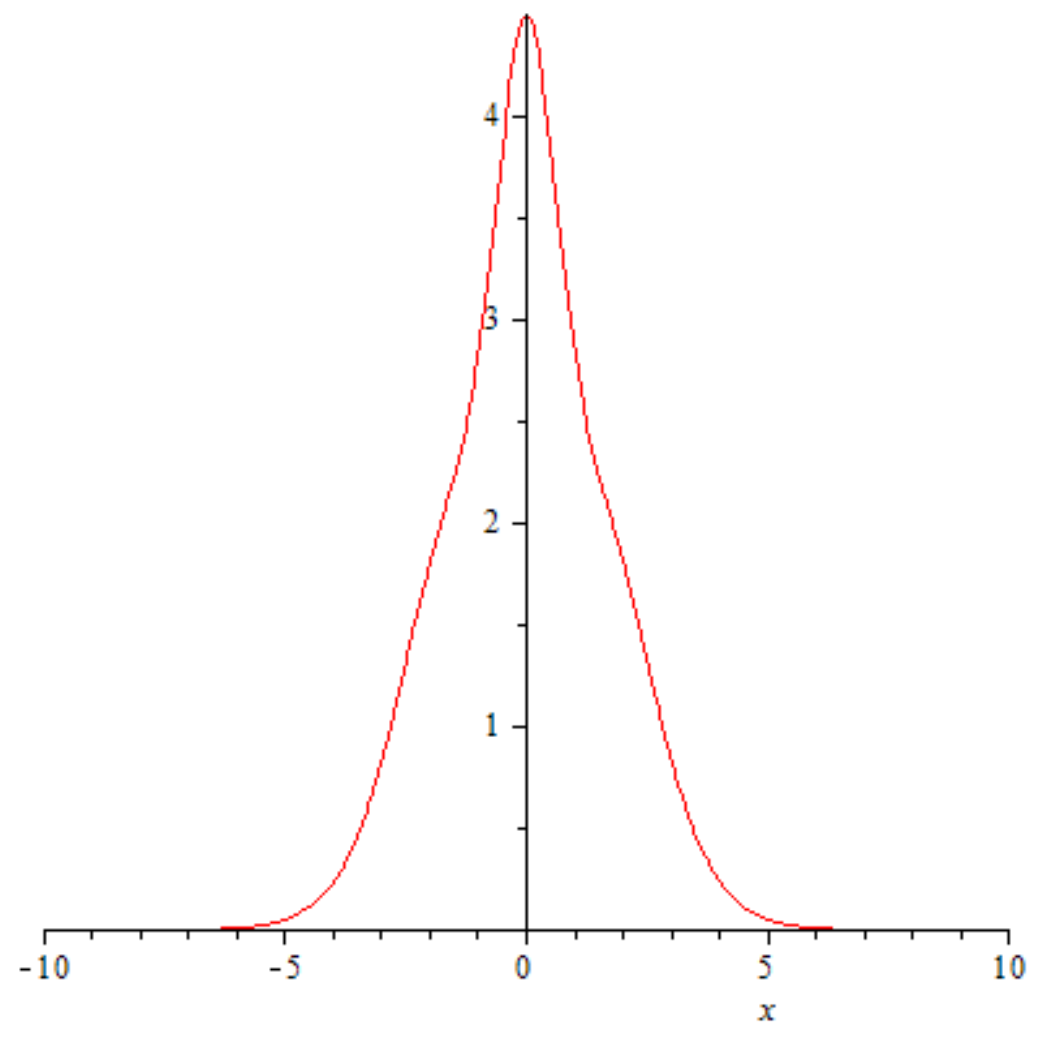}
\caption{\label{fig6} A graph of function $\left|u^{(+1,-1)}(x,t)\right|^2$ at $t=-0.2$.}
\end{center}
\end{figure}

\begin{figure}
\begin{center}
\includegraphics[width=0.6\columnwidth]{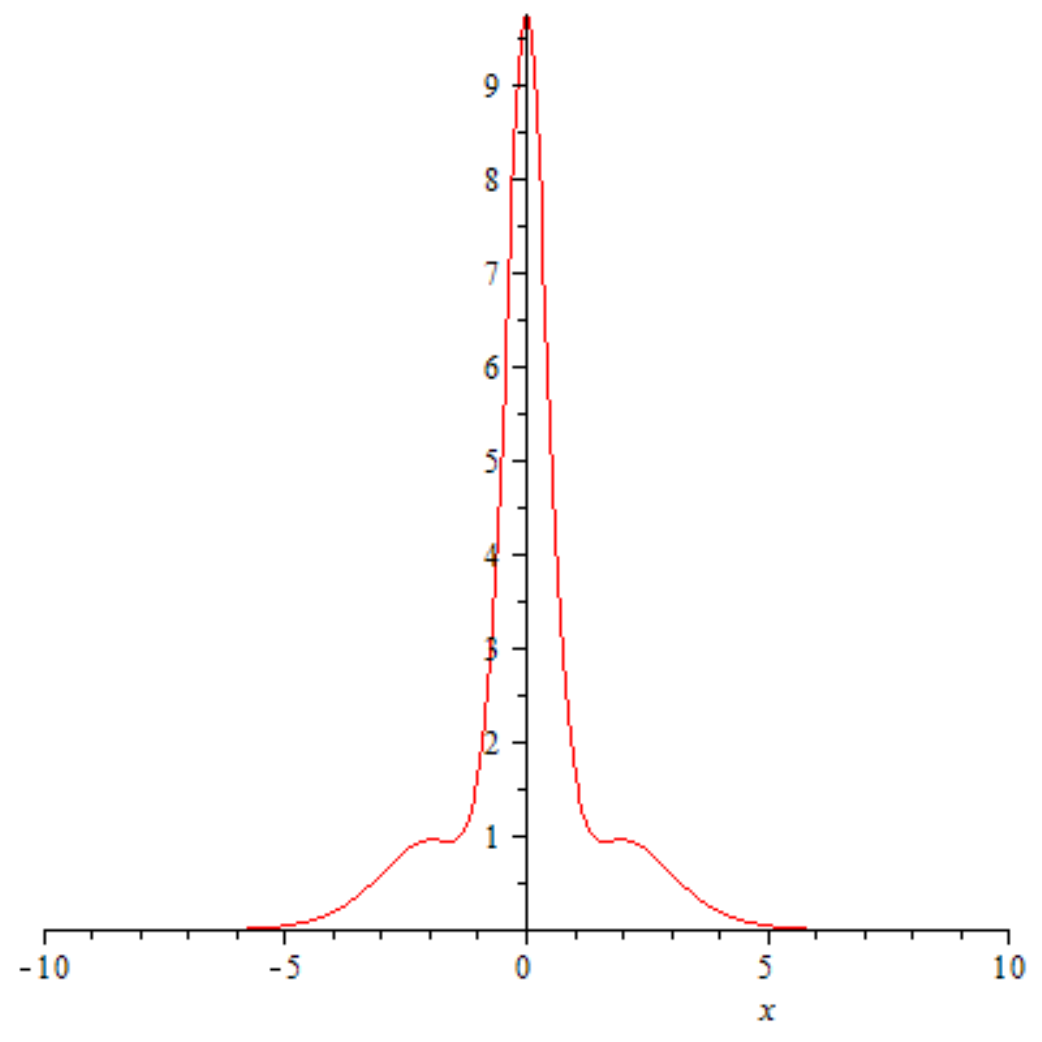}
\caption{\label{fig7} A graph of function $\left|u^{(+1,-1)}(x,t)\right|^2$ at $t=-0.1$.}
\end{center}
\end{figure}

\begin{figure}
\begin{center}
\includegraphics[width=0.6\columnwidth]{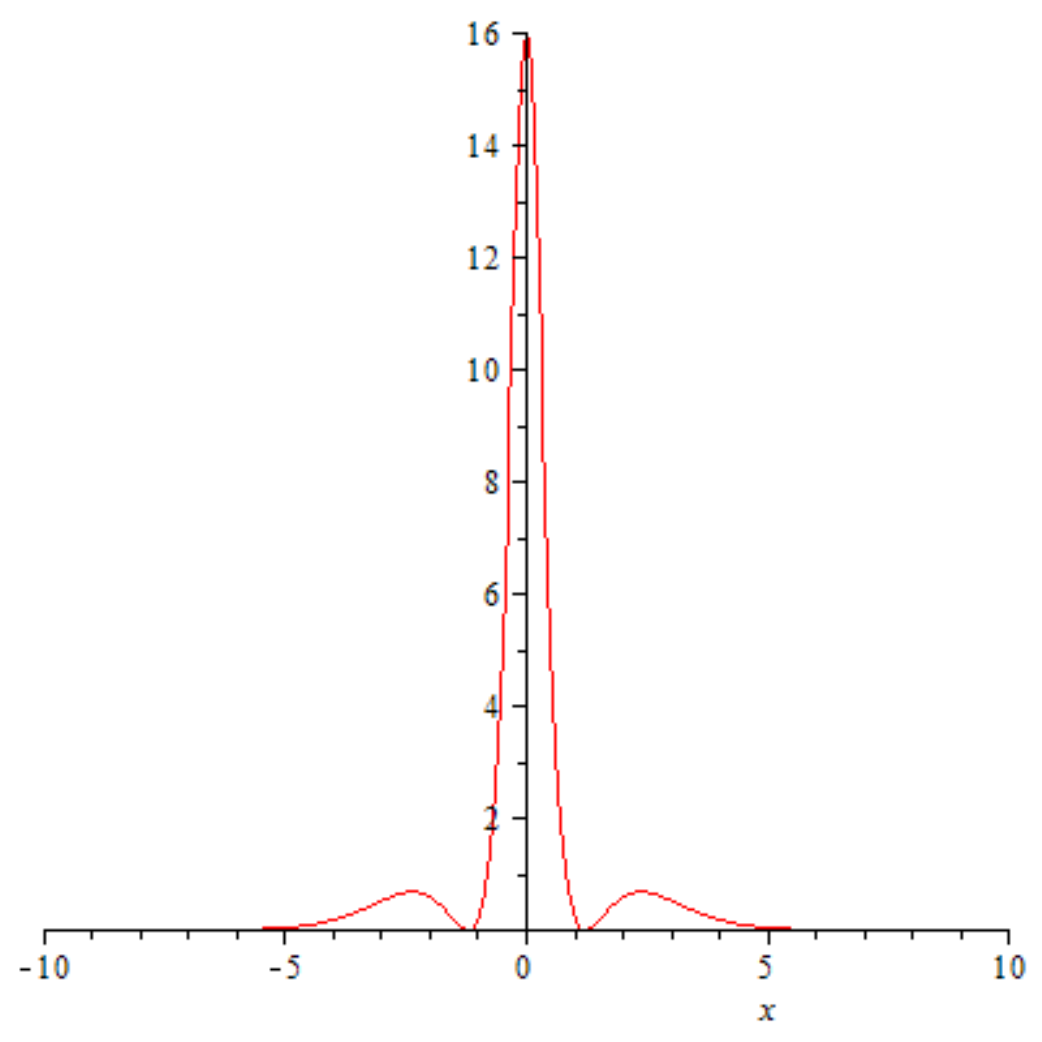}
\caption{\label{fig8} A graph of function $\left|u^{(+1,-1)}(x,t)\right|^2$ at $t=0$.}
\end{center}
\end{figure}

The amplitude of the peak changes as
\beqn \label{peak}
\left|u^{(+1,-1)}(0,t)\right|^2\sim \frac{16\beta^2}{1+64\beta^2t^2},
\enqn
reaching its maximal value at $t=0$ (Figs \ref{fig6}-\ref{fig8}).

The symmetry of the solution with respect to $t$ results in the process repeating itself for $t>0$, albeit in reverse: the high slender peak at $x=0$ is retracted, the individual solitons reemerge and continue moving in their original directions.

\begin{figure}
\begin{center}
\includegraphics[width=0.6\columnwidth]{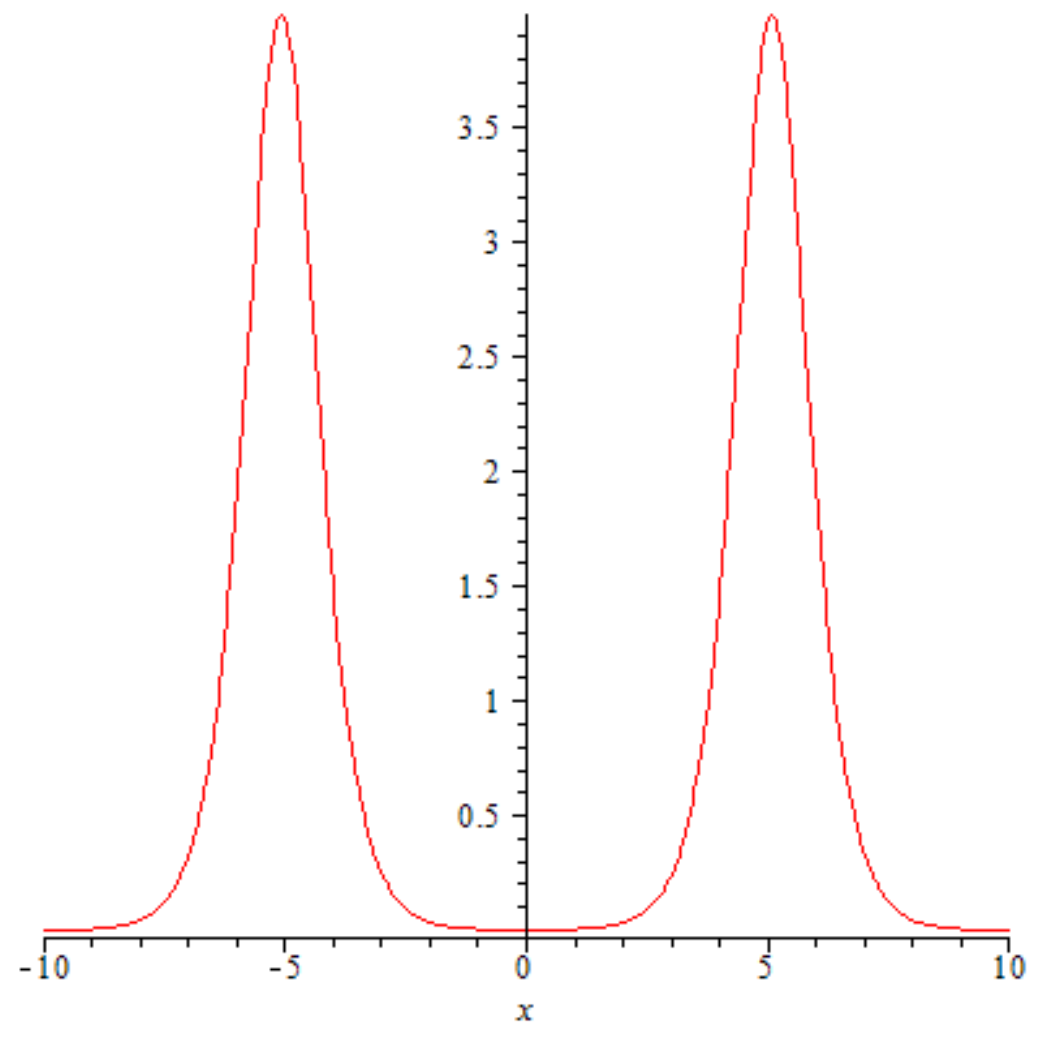}
\caption{\label{fig9} A graph of function $\left|u^{(+1,-1)}(x,t)\right|^2$ at $t=+10$.}
\end{center}
\end{figure}

Let us now take a closer look at those soliton-like solutions. Let $X=2\beta x$ and let's assume that at some time $t=t_0 < 0$ the rightmost ``soliton'' has its maximum
localized at some $X=X_0$. The symmetry of the solution implies that the left ``soliton'' has its maximum located at $X=-X_0$. This means that the derivative of \eqref{Pos-sq} with respect to $X$ should change sign at $X=\pm X_0$, $t=t_0$. This allows us to derive $t_0$ from $X_0$:
\begin{equation} \label{Pos-t0}
t_0^2=\frac{c_0^3s_0+3X_0c_0^2-2X_0^2s_0c_0\pm\sqrt{g_0}}{128 \beta^4 s_0c_0},
\end{equation}
where $c_0=\cosh(X_0)$, $s_0=\sinh(X_0)$ and
\beqn
g_0=-8s_0c_0X_0^3-c_0^2\left(15c_0^2-24\right)X_0^2+6c_0^3s_0\left(4-c_0^2\right)X_0+s_0^2c_0^4\left(c_0^2+8\right).
\enqn

For sufficiently large $X$ (or $t$) \eqref{Pos-t0} can be simplified to:
\begin{equation}
t_0=\frac{c_0}{8\beta^2}.
\label{Pos-t0big}
\end{equation}
If we the substitute this value into \eqref{Pos-sq}, we will end up with the following conclusion: for all sufficiently large $t$ the solution consists of two humps that with a very high accuracy (for $X_0=10$ the numerical analysis shows the error margin about $10^{-5}-10^{-7}$) have the form
\begin{equation}
|u|^2\sim \frac{4\beta^2}{\cosh^2(X-X_0)}.
\label{Pos-1sol}
\end{equation}

In other words, we are indeed looking at a wave whose dynamics describes two moving soliton solutions of NLS (proportional to a square of a hyperbolic secant) and their collision, which produces a short-lived rogue wave (at the time of impact $t=0$), whose maximal amplitude is more then twice higher then the amplitudes of the individual solitons (and thus exceeds even their combined height!). The peculiar nature of this rogue wave and the unusual circumstances of its inception warrants it a special name: an ``impacton'', denoting an ephemeral rogue wave, born at the point of impact of two colliding soliton waves.

But the surprises do not end here. In fact, the aforementioned individual solitons are quite remarkable even when taken by themselves. As we have seen, outside of the fact their impact begets a rogue wave, when they reemerge after the collision they again behave as two seemingly ordinary NLS solitons, keeping their shape and speed. What is unusual and keeps them apart from other known NLS solitons is the fact that their interaction produces absolutely {\em no phase shift}.

In order to show this, lets once again look at the Fig. \ref{fig1}. On the picture, we have two NLS solitons moving towards each other. Due to symmetry \eqref{Pos-chyot1}, our solution is invariant under the inversion $x \to -x$. This implies that both solitons must have must have identical phase shifts (in fact, these solitons are identical in every respect, with an notable exception of the direction of movement). On the other hand, according to \eqref{Pos-chyot2}, our solution for $t>0$ is the same as the time-inverted solution at $t<0$, which means that the collision does not produce any additional phase shifts in either soliton. Summing it all up, we have to conclude that the aforementioned interaction of leads to no phase shift.

This unusual property is characteristic of a special breed of solitons, discovered by V. Matveev and called him the ``positons'' (or ``neagtons'' -- for explanation of the difference between the two see \cite{Matveev1}).  These solutions has been extensively studied for the KdV equation \cite{Matveev2}, \cite{DGKM} (but has also been discovered for other types of equations, such as Sine-Gordon and Harry Dym equations \cite{Yurova}--\cite{Jaworski}). In particular, it has been shown that the KdV positons correspond to a relfectionless potential and are always singular, which ones again makes our solutions stand out as a rather different kettle of fish. We should also point out that also figuring out the exact nature of the solitons (namely, whether they belong to a class of positon or negaton solutions of NLS) in the impacton model lies out of scoop of this article, it would definitely be addressed in our subsequent publications.

Finally, what can we say about this solution from a point of view of ferromagnetic spin waves inside of a nanowire?.. As we have discussed in Section \ref{sec:Intro}, the individual $x$ and $y$ components of normalized magnetization $\vec m = \vec M / M_s$ are just the real and imaginary values of $u$. However, we have to keep in mind the condition we have imposed on $\vec m$ while deriving the NLS equation \eqref{NLS-1}: the function $|u|^2 \ll 1$; in order to achieve this, one should simply choose constant $\beta$ (see \eqref{Pos-2}); as we mentioned, the solution \eqref{Pos-sol} reaches its maximal value at $(x,t)=(0,0)$. According to \eqref{peak}, this value is equal to $16 \beta^2$. This implies that the uppermost value $\beta$ might theoretically reach is $\beta = 1/4$. However, since we want $\max |u|^2 \ll 1$, the value should be chosen to be even lower; for example, if we choose $\beta$ to be just ten times smaller, this would mean $\max |u(x,t)|^2 \le 10^{-2}$, which should satisfy our requirements just fine.

In conclusion, we would like to  point out that the method proposed in this article allows for a number of generalizations; in particular, it can utilized for the case of a one-dimensional multicomponent magnonic crystal with each the dynamics of each component described by the NLS \eqref{NLS} with its own coefficients. Using the binary Darboux on the background of planar wave for each of those components would produce the solutions with a number of free parameters that can be defined to ``stitch'' together the solutions from the different components, thus describing the movement of a P-breather through the entire crystal. The exact techniques for this case will be described in our next work.

\section{One last remark: But what of generalization?..}

In this article we have demonstrated that there exists a new simple analytic method designed to generate exact multi-soliton rogue wave solutions for the Nonlinear Schr\"odinger equation as well as for those equations that can be reduces to NLS -- including the Gross-Pitaevekii equation and the famous Landau-Lifshitz-Gilbert equation. The method proved to be so robust that it led us to the discovery of a completely new solution which we have called the ``impacton''. But there remains one question which till now has evaded our attention but which in our opinion fully deserved of a discussion. Of course, we are talking about the possible generalization of our approach for more general types of equations.

One of the easiest ways to generalize the NLS equation lies in a slight modification of its coefficients for one or more terms. In particular, if we add a small and seemingly innocuous assumption that two of those coefficients are {\em complex-valued} constants, we end up with a very famous equation called the complex-valued Ginzburg-Landau equation (CGLE):
\beq \label{Ginzburg-Landau}
\frac{\partial \psi}{\partial t} = \alpha \frac{\partial^2 \psi}{\partial x^2} + \left(\beta |\psi|^2-\sigma\right) \psi,
\enq
where $\psi=\psi(x,t)$, $\sigma \in \R$ and $\alpha$, $\beta$ are two complex-valued constants:
\beqn
\alpha = c_1 + i c_2, \qquad \beta = c_3 + i c_4, \qquad c_i \in \R.
\enqn
Similar to NLS, the CGLE arises in a multitude of physical problems; it naturally appears in such distinct areas as that of superconductivity, of the Bose-Einstein condensation and in the descriptions of the second-order phase transitions \cite{AK}.\footnote{In addition, in the special case of $c_2 = c_4 =0$, the CGLE becomes an ordinary Ginzburg-Landau equation, useful for various convection problems, such as the Rayleigh-B\'enard convection, the Taylor-Couette flow etc.}

Of course, such a splendid slew of applications did not fail to motivate the mathematicians to perform a very thorough study of \eqref{Ginzburg-Landau}, which in turn led them to a very sad but irrefutable conclusion: the Ginzburg-Landau equation proved to be generally non-integrable. There exists one obvious exception to this rule which arises when \eqref{Ginzburg-Landau} gets reduced to NLS (in fact, when $c_1=c_3=0$, $\sigma=iB^2$, and $B^2 \in \R$, the equation \eqref{Ginzburg-Landau} turns into a Gross-Pitaevskii equation), but unfortunately the majority of all physically interesting applications produce the non-integrable versions of \eqref{Ginzburg-Landau}. With that being said, the non-integrability of a non-linear differential equation does not necessarily imply that it is actually impossible to find {\em some} classes of exact solutions, nor does it preclude us from discovering the analytic means to {\em construct} such solutions. For example, it has been demonstrated by Tajiri in \cite{Tajiri} that the $(1+2)$-dimensional complex-valued nonlinear cubic Klein-Gordon equation permits for infinitely many soliton solutions that can be explicitly constructed via a simple change of variables. Thus, it is not actually unreasonable to ask ourselves: can there exist some hitherto unknown classes of exact analytic solutions for \eqref{Ginzburg-Landau}? We believe that the answer here shall be affirmative. In fact, we would like to argue that CGLE \eqref{Ginzburg-Landau} might permit for an infinitely large class of such solutions and that those solutions would probably be nothing else but some sort of deformed multi-soliton rogue wave solutions of NLS!..

Allow us to explain. First of all, the very fact of non-integrability of CGLE makes the method developed in this article virtually useless: the equation \eqref{Ginzburg-Landau} has no discernible Lax pair, nor does it allow for a kind of symmetries that lead to the Darboux transformation. But, curiously enough, this equation might be tackled using the very method by Dubard and Matveev \cite{Dubard_Matveev} which we have been aiming to surpass throughout this article! The reason for it is simple: their method of construction of multi-soliton rogue wave solutions {\em does not rely on integrability of NLS}. Instead, it is the relationship between the NLS and the non-stationary Schr\"odinger equation that is a key here. In order to illustrate this, let us first choose (for the sake of simplicity) $\sigma=0$ and then introduce two new variables:
$$
z=\frac{t}{\alpha^*},\qquad y=\frac{x}{|\alpha|},
$$
a new complex constant $\mu=\alpha^*\beta$ and a real-valued potential $V=\psi^*\psi$.  The equation \eqref{Ginzburg-Landau} can then be formally rewritten as a ``non-stationary Schr\"odinger equation'' with a complex (not real!) time $z$ and the unchanged space variable $y$:
\beq \label{CNSEq}
\frac{\partial \psi}{\partial z} = \frac{\partial^2 \psi}{\partial y^2} + \mu V \psi.
\enq
In fact, if $z$ is purely imaginary, and $\mu$ is real, \eqref{CNSEq} turns into an ordinary non-stationary Schr\"odinger equation, whereas \eqref{Ginzburg-Landau} turns into Gross-Pitaevskii equation (or, in other words, into NLS \eqref{NLS}). It is this very property that serves as a starting point for the reasoning developed in \cite{Dubard_Matveev}. In short, the reasoning consists of two lemmas: (i) that the non-stationary Schr\"odinger equation permits for arbitrarily many iterations of the Darboux transformation and (ii) for those solutions produced after N consecutive Darboux transformations over zero background there exists such a set of parameters that the new solution $|\psi|^2$ is {\em linearly dependent} on N-th iteration of the complex-valued potential $V$. Dubard and Matveev then used these two lemmas to iteratively construct the whole class of solutions satisfying both (i) and (ii). And those solutions were none other than the multi-soliton rogue waves!

Using a similar logic, the problem posed in this section can be reformulated as follows: does there exist solutions for the non-stationary Schr\"odinger equation with a {\em complex} time \eqref{CNSEq} that respects both (i) and (ii) and that also keep the potential $V$ real-valued? The proposition (i) is obviously true: indeed, if $\{\psi_1,...,\psi_n,\,\psi\}$ are the solutions of \eqref{CNSEq} with $V=0$, then
$$
\psi^{(n)}=\frac{W_2}{W_1},\qquad V^{(N)}=\frac{2}{\mu} \frac{\partial^2 \log W_1}{\partial y^2},
$$
satisfy the equation
$$
\frac{\partial \psi^{(n)}}{\partial z} = \frac{\partial^2 \psi^{(n)}}{\partial y^2} + \mu V^{(n)} \psi^{(n)},
$$
where by $W_{1,2}$ we understand the Wro\'nskians
$$
W_2=W(\psi_1,...,\psi_n,\,\psi),\qquad W_1=W(\psi_1,...,\psi_n)={\rm det}\left(\partial^{i-1}_y\psi_j\right),\qquad   i,j=1..n.
$$
Hence, in order to prove that our hypothesis is correct it remains to show that there exist such $\{\psi_k\}_{k=0}^{N-1}$, that $V^{(N)}$ would be a real-valued function, and that for such functions the proposition (ii) will hold, i.e. that there exist such constant $\gamma, \delta \in \R$, that
$$
\left|\frac{W_2}{W_1}\right|^2=\gamma~\frac{\partial^2 \log W_1}{\partial y^2} + \delta.
$$

As with the NLS case, we strongly suspect that such functions both exist and have the characteristic properties of multi-soliton rogue waves (albeit appropriately deformed). But the proof of these remaining statements lies too far outside of the scope of this article, and will therefore have to be addressed elsewhere.

\section{Acknowledgments} \label{sec:Acknowledgements}

This article has been written as a part of a strategic initiative ``Functional Magnetic Materials for Biomedical and Energy Relevant Applications'' (FunMagMa) at the Immanuel Kant Baltic Federal University. Valerian Yurov would like to thank his colleagues from the FunMagMa project for their interest in the project and especially for the wonderful seminars, that eventually led us to the crucial idea lying behind this article.





\end{document}